\documentclass[12pt,usenames,dvipsnames]{article}

\usepackage{latexsym}
\usepackage{amssymb,amsfonts,amsmath}
\usepackage{graphicx} 
\usepackage{indentfirst}
\usepackage{bbm}
\usepackage{amssymb}
\usepackage{verbatim}
\usepackage{amsmath, amsthm,amssymb}
\usepackage{mathrsfs}
\usepackage{hyperref}
\usepackage{amsfonts}
\usepackage{dsfont}
\usepackage{cite}
\usepackage{xcolor}
\usepackage{enumerate}
\usepackage{cleveref}
\usepackage{cjhebrew}
\usepackage{arabtex}

\parskip=\medskipamount

\newcommand {\cD}{{\cal D}}

\newcommand {\cF}{{\cal F}}

\newcommand {\cI}{{\cal I}}

\newcommand {\cK}{{\cal K}}

\newcommand {\cM}{{\cal M}}
\newcommand {\cN}{{\cal N}}

\newcommand {\cR}{{\cal R}}

\newcommand {\cT}{{\cal T}}
\newcommand {\cU}{{\cal U}}

\def\a{\alpha}

\def\b{\beta}

\def\d{\delta}

\def\g{\gamma}

\def\l{\lambda}
\def\m{\mu}

\def\o{\omega}

\def\q{\theta}
\def\r{\rho}
\def\s{\sigma}
\def\t{\tau}

\def\x{\xi}

\def\F{\Phi}

\def\L{\Lambda}
\def\O{\Omega}
\def\P{\Pi}

\def\rd{{\rm d}}
\def\ri{{\rm i}}
\def\re{{\rm e}}

\def\N{{\cal N}}

\newcommand{\ve}{\varepsilon}                            

\newcommand{\ab}{{\a\b}}

\newcommand{\pa}{\partial}                           
\newcommand{\hf}{\frac12}

\newcommand{\be}{\begin{equation}}
\newcommand{\ee}{\end{equation}}
\newcommand{\bea}{\begin{eqnarray}}
\newcommand{\eea}{\end{eqnarray}}
\newcommand{\non}{\nonumber}

\newcommand{\Iu}{{\underline{I}}}

\def\dt#1{{\buildrel {\hbox{\LARGE .}} \over {#1}}}    

\newcommand{\bm}[1]{\mbox{\boldmath$#1$}}

\def\double #1{#1{\hbox{\kern-2pt $#1$}}}

\newcommand{\OI}{{\overline{I}}}
\newcommand{\OJ}{{\overline{J}}}
\newcommand{\OK}{{\overline{K}}}
\newcommand{\OL}{{\overline{L}}}
\newcommand{\OM}{{\overline{M}}}

\newcommand{\UI}{{\underline{I}}}
\newcommand{\UJ}{{\underline{J}}}
\newcommand{\UK}{{\underline{K}}}
\newcommand{\UL}{{\underline{L}}}
\newcommand{\UM}{{\underline{M}}}

\newif\ifdtup

\newcommand{\bsubeq}{\begin{subequations}}
\newcommand{\esubeq}{\end{subequations}}

\numberwithin{equation}{section}

\newcommand{\sSU}{\mathsf{SU}}
\newcommand{\sSL}{\mathsf{SL}}

\newcommand{\sSO}{\mathsf{SO}}

\newcommand{\sU}{\mathsf{U}}

\newcommand{\sOSp}{\mathsf{OSp}}

\newcommand{\id}{\mathds{1}}

\newcommand{\uI}{{\underline{I}}}

\begin{document}
	
\begin{titlepage}
	\begin{flushright}
	June, 2025
	\end{flushright}
	\vspace{5mm}
	
	\begin{center}
		{\Large \bf 
			New superparticle models in AdS superspaces
		}
	\end{center}

\begin{center}
		{\bf Nowar E. Koning
        and Emmanouil S. N. Raptakis} \\
		\vspace{5mm}
		
		\footnotesize{ 
			{\it Department of Physics M013, The University of Western Australia\\
				35 Stirling Highway, Perth W.A. 6009, Australia}}  
		~\\
		\vspace{2mm}
		~\\
		Email: \texttt{nowar.koning@research.uwa.edu.au,
            emmanouil.raptakis@uwa.edu.au }\\
		\vspace{2mm}
	\end{center}
	
\begin{abstract}
		\baselineskip=14pt
Recently, new superparticle models have been proposed in the $\mathcal{N}$-extended four and five-dimensional anti-de Sitter (AdS) superspaces, AdS$^{4|4\mathcal{N}}$ and AdS$^{5|8\mathcal{N}}$, making use of a unique quadratic deformation to the AdS supersymmetric interval. 
In this paper we extend these considerations to the three and two-dimensional cases, and propose new two-derivative models for superparticles propagating in these AdS superspaces. 
\end{abstract}
\vspace{5mm}
	
	\vfill
	
	\vfill
\end{titlepage}

\newpage
\renewcommand{\thefootnote}{\arabic{footnote}}
\setcounter{footnote}{0}

\tableofcontents{}
\vspace{1cm}
\bigskip\hrule

\allowdisplaybreaks

\section{Introduction}

Anti-de Sitter supergeometries have been a topic of interest in the literature since their introduction in the early years of supersymmetry by Keck \cite{Keck} and Zumino \cite{Zumino}. 
The first to appear was the four-dimensional $\N=1$ AdS superspace, AdS$^{4|4}$, and the comprehensive analysis of general supermultiplets on $\text{AdS}^{4|4}$ was given in a paper by Ivanov and Sorin \cite{IS}, arguably one of the most important works on AdS supersymmetry.\footnote{See also \cite{IS2} by the same authors.}
In general, there exist two main approaches when considering such supergeometries: (i) a supergravity-inspired framework; and (ii) an appropriate embedding formalism. 
In the former, AdS backgrounds are singled out by the requirement that the torsion and curvature tensors are Lorentz-invariant and covariantly constant, see, e.g., \cite{KLTM, KTM2, KKR}, and references therein.
The latter stems from supersymmetric extensions to the standard embedding realisation of AdS$_d$ as a hypersurface in $\mathbb{R}^{d-1\,,2}$ 
\begin{align}
    - (Z^{0})^{2} + (Z^{1})^{2} + \cdots + (Z^{d-1})^{2} - (Z^{d})^{2} = -\ell^{2} = \text{const}.
\end{align}
Such extensions have been developed in the three, four, and five-dimensional cases \cite{KTM, KT, KKR, KK}.\footnote{For a review of superembedding approaches, see \cite{DS, BDS}.}  

The $\N$-extended conformally flat AdS superspaces in three dimensions can be defined from a group-theoretic point of view as 
\begin{align} \label{3d ads coset}
    \text{AdS}_{(3|p\,,q)} = \frac{\sOSp(p|2;\mathbb{R})\times\sOSp(q|2;\mathbb{R})}{\sSL(2;\mathbb{R})\times\sSO(p)\times\sSO(q)}\,,
\end{align}
see, e.g. \cite{BILS}. 
They were introduced in the supergravity setting in \cite{KLTM} as backgrounds of the $\N$-extended off-shell conformal supergravity in three dimensions \cite{HIPT, KLTM2}.
The conformal flatness of AdS$_{(3|p\,,q)}$ was argued in \cite{BILS} and a conformally flat realisation was derived in \cite{KLTM}, based on the use of stereographic coordinates. 
The study of AdS$_{(3|p\,,q)}$ as the homogeneous space \eqref{3d ads coset} was carried out in \cite{KT}.

In recent years, supersymmetric theories in $\text{AdS}_2$ have attracted much interest, see, e.g. \cite{BHT} and references therein.
These studies have largely occurred within a component framework. As an alternate formulation, the two-dimensional $\cN$-extended AdS superspaces were constructed in \cite{KR} by making use of the $\sSO(p) \times \sSO(q)$ superspace formulations for conformal supergravity developed in the same work.
These $\cN$-extended AdS superspaces are characterised by isometry supergroups of $\sOSp(\cN|2;\mathbb{R})$.
There are potentially other AdS superspaces in two dimensions, characterised by different isometry supergroups. These isometry supergroups were deduced in \cite{GST}. 

Superparticle models with spacetime supersymmetry were first introduced in the late 1970s by Casalbuoni \cite{C}, and were soon after reformulated by Brink and Schwarz \cite{BS}. They have by now become a textbook subject \cite{PW}.  
Recently, both approaches\footnote{In addition, (super)twistor techniques have been used extensively in the literature to describe (super)particles in AdS, see \cite{CGKRZ, CRZ, CKR,BLPS,Z,Cm,Cm2,ABGT,ABGT2,U,U2}.} to AdS superspace have been used to derive new models for superparticles propagating in the four-dimensional $\N$-extended AdS superspace, AdS$^{4|4\N}$ \cite{KKR2}.\footnote{They have also been extended to the five-dimensional case, see \cite{KK}.}
The key feature of these new superparticle models is that, 
in contrast to the standard Brink-Schwarz superparticle, which is derived from an interval of the form 
\begin{align}
    \rd s^{2} = \eta_{ab}E^{a}E^{b}\,, \qquad a = 0\,,1\,,2\,,3\,,
\end{align}
where $\eta_{ab} = \text{diag}(-1\,,1\,,1\,,1)$ is the Minkowski metric in four dimensions, $E^{a}= \rd z^{M}E_{M}{}^{a}$ is the vector component of the supervielbein, and $z^{M}$ are local coordinates, they make use of an AdS-specific deformation to the supersymmetric interval. 
On the one hand, such a deformation is available in the supergravity setting due to the presence of a nowhere-vanishing dimension-1 torsion superfield.  
On the other hand, one can make use of the unique (two-point) invariants of the AdS supergroup to construct the deformed interval in the embedding formalism. 
The present paper is devoted to extending these considerations to the three and two-dimensional AdS superspaces, and deriving both the deformed supersymmetric intervals and their corresponding superparticle models.

This paper is organised as follows. 
In section \ref{section 2} we introduce the deformed supersymmetric interval in the three-dimensional $(p\,,q)$ AdS superspace, and use it to derive a new superparticle model. We then rederive this model from the embedding formalism and prove their equivalence. 
In section \ref{section 3}, we extend these considerations to derive the deformed interval and corresponding model in the two-dimensional AdS supergeometry.
In section \ref{discussion}, we provide concluding comments and some generalisations of the three-dimensional case.
The main body of this paper is accompanied by two technical appendices. Appendix \ref{appendix a} outlines the necessary material for conformal supergravity in three dimensions, including our spinor conventions. 
Appendix \ref{appendix b} reviews the conformal superspace approach to conformal $(p\,,q)$ supergravity in two dimensions.

\section{AdS superspaces in three dimensions} \label{section 2}

In this section we introduce the deformed supersymmetric interval in the three-dimensional AdS supergeometry, first in the supergravity setting, and then from the embedding formalism.
Making use of this deformed interval, we derive the corresponding superparticle model. 
Our conventions for conformal supergravity in three dimensions are described in appendix \ref{appendix a}.

\subsection{$(p\,,q)$ AdS superspace}

We consider a three-dimensional curved superspace $\cM^{3|2\N}$, parametrised by local bosonic $(x^{m})$ and fermionic $(\q_{\tt I}^{\m})$ coordinates 
\begin{align}
    z^{M} = (x^{m}\,, \q_{\tt I}^{\m})\,, \qquad m = 0\,,1\,,2\,, \quad \m = 1\,,2\,, \quad {\tt I} = 1\,, \ldots\,, \N\,.
\end{align}
For the $\N=1$ case, the superspace geometry is controlled by the torsion superfields 
\begin{align}
    C_{\a\b\g} = C_{(\a\b\g)}\,, \qquad S\,,
\end{align}
whereas for $\N>1$ we have
\begin{align}
    X^{IJKL} = X^{[IJKL]}\,, \qquad C_{a}{}^{IJ} = C_{a}{}^{[IJ]}\,, \qquad S^{IJ} = S^{(IJ)}\,.
\end{align}
We emphasise that for $\N<4$ the super-Cotton tensor $X^{IJKL}$ vanishes identically, $X^{IJKL} = 0\,.$ 
As pointed out in the introduction, the AdS supergeometry can be singled out by the requirement that these torsion superfields are Lorentz-invariant and covariantly constant. 
This implies
\bsubeq \label{3d ads constraints}
\begin{align} 
\N &= 1: \qquad C_{\a\b\g} = 0\,, \qquad \cD_{A}S = 0\,,
\\
\N&>1: \qquad     C_{a}{}^{IJ} = 0\,, \qquad \cD_{A} S^{JK} = 0\,, \qquad \cD_{A}X^{JKLM} =0\,,  
\end{align}
\esubeq
where $\cD_{A} = (\cD_{a}\,, \cD_{\a}^{I})$ are the covariant derivatives.
Keeping these constraints in mind, the covariant derivatives obey the algebra
\bsubeq
\begin{align}
    \{\cD_{\a}^{I}\,,\cD_{\b}^{J}\} &= -2\ri\d^{IJ}\cD_{\ab} - 4\ri S^{IJ}\cM_{\ab} + \ri\ve_{\ab}(X^{IJKL}-4S^{K[I}\d^{J]L})\N_{KL}\,,
    \\
    [\cD_{a}\,,\cD_{\a}^{I}] &= S^{I}{}_{J}(\g_{a})_{\a}{}^{\b}\cD_{\b}^{J}\,,
    \\
    [\cD_{a}\,,\cD_{b}] &= -4S^{2}\cM_{ab}\,,
\end{align}
\esubeq
where $S^{2} := \frac{1}{\N}S^{IJ}S_{IJ}\,.$
This is the most general $\N$-extended AdS supergeometry of \cite{KLTM}. 

The superfield $S^{IJ}$ and super-Cotton tensor $X^{IJKL}$ satisfy non-trivial integrability conditions.
These are
\bsubeq
\begin{align} \label{s props}
    S^{IK}S_{K}{}^{J} = S^{2}\d^{IJ}\,,
\end{align}
and
\begin{align} \label{x const}
   X_{N}{}^{IJ[K}X^{LPQ]N} = 0 \implies X^{KLM[I}X^{J]}{}_{KLM} = 0\,. 
\end{align}
\esubeq
It follows from \eqref{s props} that, by a local $\sSO(\N)$ transformation, $S^{IJ}$ can be brought to the form 
\begin{align} \label{3d diag}
S^{IJ} = S ~ \text{diag}(\overbrace{+1\,, \cdots \,, +1}^{p} \,, \overbrace{-1\,, \cdots \,, -1}^{q = \N-p})\,, 
\end{align}
resulting in an unbroken local group $\sSO(p)\times \sSO(q)\,.$ 
The corresponding supergeometry is referred to as AdS$_{(3|p\,,q)}$.
Further, it was shown in \cite{KLTM} that a non-zero super-Cotton tensor $X^{IJKL}$ is only compatible with positive-definite $S^{IJ}\,.$ 
In other words, all three-dimensional $(p\,,q)$ AdS superspaces with $q >0$ have vanishing $X^{IJKL}\,.$

For the remainder of this section we will only consider the $q>0 \implies X^{IJKL} = 0$ case. This setup will be revisited in section \ref{discussion}.

\subsection{Conformally flat frame}

Bringing the superfield $S^{IJ}$ to the diagonal form \eqref{3d diag} is a special gauge choice which was made use of in \cite{KLTM}. 
It is often useful to relax this gauge condition, and unless otherwise stated we will assume that $S^{IJ}$ is not in the diagonal form \eqref{3d diag}. 
The advantage of doing so is that one can preserve local $\sSO(\N)$ freedom and, as a result, have a frame in which the covariant derivatives are related to those of the 3D Minkowski superspace by a super-Weyl transformation \eqref{cd trans}.\footnote{Indeed, as described in \cite{KKR2}, the gauge choice \eqref{3d diag} is not compatible with the frame \eqref{3d cflat derivatives}.}

In this case, the AdS covariant derivatives $\cD_{A}$ in three dimensions look like
\bsubeq \label{3d cflat derivatives}
\begin{align}
    \cD_{\a}^{I} &= \re^{\frac{1}{2}\s}\left\{ D_{\a}^{I} + (D^{\b I}\s)\cM_{\ab} + (D_{\a J}\s)\N^{IJ} \right\}\,,
    \\
    \cD_{a} &= \re^{\s}\bigg\{\partial_{a} + \frac{\ri}{2}(\g_{a})^{\ab}(D^{K}_{(\a}\s)D_{\b)K} + \ve_{abc}(\partial^{b}\s)\cM^{c} + \frac{\ri}{16}(\g_{a})^{\ab}([D^{[K}_{(\a}\,,D^{L]}_{\b)}]\s)\N_{KL}
    \\
    \notag 
    & ~~~~~~~\qquad - \frac{\ri}{8}(\g_{a})^{\ab}(D_{K}^{\r}\s)(D^{K}_{\r}\s)\cM_{\ab} + \frac{3\ri}{8}(\g_{a})^{\ab}(D^{[K}_{(\a}\s)(D_{\b)}^{L]}\s)\N_{KL}
    \bigg\}\,, 
\end{align}
\esubeq
where $D_{\a}^{I}$ is the flat 3D spinor derivative, defined as
\begin{align}
    D_{\a}^{I} = \partial_{\a}^{I} + \ri(\g^{c})_{\ab}\q^{\b I}\partial_{c}\,, \qquad \{D_{\a}^{I}\,,D_{\b}^{J}\} = 2\ri\d^{IJ}(\g^{c})_{\ab}\partial_{c}\,,
\end{align}
and $\s = \bar{\s}$ is the super-Weyl parameter, enjoying the constraint 
\bsubeq  \label{3d ads constraint}
\begin{align}
\N&=1: \qquad \partial_{(\ab}D_{\g)}\re^{\s} = 0\,,
\\
\N&>1: \qquad D_{(\a}^{[I}D_{\b )}^{J]}\re^{\s} = 0\,.    
\end{align}
\esubeq
Such a frame will be called conformally flat. 
We point out that the expressions \eqref{3d cflat derivatives} are also valid for $\N=1$. In this case, the terms containing the $\sSO(\N)$ generators $\N_{IJ}$ disappear, see eq. \eqref{n=1 sw cd}. 
Solutions to \eqref{3d ads constraint} were derived in \cite{KLTM, KT} making use of stereographic and Poincar\'e coordinates, respectively. 

It follows from \eqref{3d cflat derivatives} that the components of the supervielbein $E^{A} = (E^{a}\,, E^{\a}_{I})$ are given by
\bsubeq \label{cflat 3d vb}
\begin{align}
    E^{a} &= \re^{-\s}\P^{a}\,, \qquad \P^{a} = \rd x^{a} + \ri\q_{I}^{\a}(\g^{a})_{\ab}\rd\q_{I}^{\b}\,, \label{cflat 3d vector}
    \\
    E_{I}^{\a} &= \re^{-\frac{1}{2}\s}\left\{ \rd\q_{I}^{\a} - \frac{\ri}{2}(D_{\b I}\s)\P^{\ab} \right\}\,. 
\end{align}
\esubeq
Here, $\P^{a}$ is the three-dimensional analogue of the Volkov-Akulov supersymmetric one-form.

\subsection{Interval deformation for $p\geq q>0$}

We will now use the above considerations to derive a quadratic deformation to the supersymmetric interval in AdS$_{(3|p\,,q)}$.

The standard supersymmetric interval is constructed from the vector component of the supervielbein, and is given by
\begin{align} \label{3d interval}
    \rd s^{2} = \eta_{ab}E^{a}E^{b}\,, \qquad \eta_{ab} = \text{diag}(-1\,,+1\,,+1)\,.
\end{align}
Making use of the torsion superfield $S^{IJ}$ and the spinor components of the supervielbein $E^{\a}_{I}\,,$ we can introduce a one-parameter deformation as follows 
\begin{align} \label{3d def}
    \rd s^{2} = E^{A}\eta_{AB}E^{B} = \eta_{ab}E^{a}E^{b} + \frac{\ri\o}{S^{2}}S^{IJ}\ve_{\ab}E_{I}^{\a}E_{J}^{\b}\,,
\end{align}
for a real, dimensionless parameter $\o$.
The supermatrix $\eta_{AB}$ is defined as 
\begin{align} \label{supermetric}
    (\eta_{AB}) = \left(\begin{array}{c||c}
        ~\eta_{ab}~ & 0 \\
        \hline \hline
        ~0~ & \frac{\ri\o}{S^{2}}S^{IJ}\ve_{\ab}
    \end{array}\right)\,. 
\end{align}
We emphasise that the deformation \eqref{3d def} is invariant under the three-dimensional AdS superisometries.

Given the deformation \eqref{3d def}, we can introduce a new model for a superparticle propagating in AdS$_{(3|p\,,q)}$, defined by 
\begin{align} \label{3d model}
    S = \frac{1}{2}\int \rd\t \frak{e}^{-1}\left\{\dt{E}{}^{A}\eta_{AB}\dt{E}{}^{B} - (\frak{e}m)^{2}\right\}\,, \qquad \dt{E}{}^{A} := \frac{\rd z^{M}}{\rd \t} E_{M}{}^{A}\,,
\end{align}
where $\frak{e}$ is the einbein, $m$ is the mass, and $\t$ is the evolution parameter. 
In the conformally flat frame, where the components of the supervielbein are given by the expressions \eqref{cflat 3d vb}, we find 
\begin{align}
    \dt{E}{}^{A}\eta_{AB}\dt{E}{}^{B} &= \re^{-2\s}\dt{\P}{}^{2} + \frac{\ri\o}{S^{2}}\re^{-\s}S^{IJ}\left\{ \dot{\q}_{IJ} - \ri(D_{\a I}\s)\dt{\P}{}^{\ab}\dot{\q}_{\b J} + \frac{1}{4}(D^{\a}_{I}\s)(D_{\a J}\s)\dt{\P}{}^{2}
   \right\}\,, 
\end{align}
where $\dt{\P}{}^{a} = \dot{x}{}^{a} + \ri\q_{I}^{\a}(\g^{a})_{\ab}\dot{\q}{}_{I}^{\b}\,.$
Further, in the diagonal frame \eqref{3d diag}, we have
\begin{align}
    \dt{E}{}^{A}\eta_{AB}\dt{E}{}^{B} = \eta_{ab}\dt{E}{}^{a}\dt{E}{}^{b} + \frac{\ri\o}{S}\ve_{\ab}\left\{ \dt{E}{}_{\bar{I}}^{\a}\dt{E}{}_{\bar{I}}^{\b} - \dt{E}{}_{\uI}^{\a}\dt{E}{}_{\uI}^{\b} \right\}\,, 
\end{align}
with $\OI = 1\,, \ldots\,, p$ and $\UI = p+1\,,\ldots\,,\N\,.$
We point out that for $\o = 0$, we recover the standard superparticle propagating in AdS superspace.
As a result, the model \eqref{3d model} is the three-dimensional analogue of those proposed in \cite{KKR2} and \cite{KK}.

It is of interest to see how the above two-derivative structures arise in the embedding formalism for AdS$_{(3|p\,,q)}$ derived in \cite{KT}. 
In this formalism, the most general superparticle model quadratic in derivatives of the evolution parameter $\t$ is given by the following 
\begin{align} \label{3d embed model}
    S = -\frac{1}{2}\int \rd \t \frak{e}^{-1} \left\{\frac{1}{2} \text{Str}(\dot{\bm{Z}}\dot{\bm{Z}}) + \bigg(\a- \frac{1}{4}\bigg)\big(\text{Str}(\dot{\bm{X}}\dot{\bm{X}}) + \text{Str}(\dot{\bm{Y}}\dot{\bm{Y}})\big) + (\frak{e}m^{2})\right\}\,, 
\end{align}
with $\a \in \mathbb{R}$ and where $\bm{X}\,, \bm{Y}\,,$ and $\bm{Z}$ are bi-supertwistors of AdS$_{(3|p\,,q)}\,,$ see \cite{KT} for more details on the bi-supertwistor construction. 
The coefficients of each structure are chosen to satisfy the following two conditions: (i) the model coincides with the bosonic one when the Grassmann variables are switched off; and (ii) the $\a = 0$ case recovers the non-deformed superparticle model. 
Indeed, it was shown in \cite{KT} that the following combination 
\begin{align}
    \rd s^{2} = -\frac{1}{2}\text{Str}(\rd \bm{Z} \rd\bm{Z}) + \frac{1}{4}(\text{Str}(\rd\bm{X}\rd\bm{X}) + \text{Str}(\rd\bm{Y}\rd\bm{Y})),
\end{align}
yields the two-point function \eqref{3d interval}.  
Furthermore, in the Poincar\'e-like\footnote{Strictly speaking, reference \cite{KT} introduced two coordinate systems for AdS$_{(3|p\,,q)}$ based on the use of Poincar\'e coordinates. One provided a conformally flat realisation, see eq. \eqref{3d cflat derivatives}, and the other corresponded to the diagonal frame \eqref{3d diag}. We refer to the latter as Poincar\'e-like coordinates.} coordinate system developed for the embedding formalism in \cite{KT}, which makes use of the gauge choice \eqref{3d diag}, the additional structures present in \eqref{3d embed model} take the form
\bsubeq
\begin{align}
    \text{Str}(\dot{\bm{X}}\dot{\bm{X}}) &= 4\ri z^{-2}\left(\dot{\q}^{+}_{\OI}\q^{-}_{\OI}(\dot{z}+ \ri\dot{\q}^{+}_{\OJ}\q^{-}_{\OJ}) + \dot{\q}^{-}_{\OI}\q^{-}_{\OI}(\dot{u}^{++} - \ri\dot{\q}^{+}_{\OJ}\q^{+}_{\OJ})
  \right) + 4\ri z^{-1}\dot{\q}^{-}_{\OI}\dot{\q}^{+}_{\OI}\,, 
  \\
  \text{Str}(\dot{\bm{Y}}\dot{\bm{Y}}) &= 4\ri z^{-2} \left( 
  -\dot{\q}^{-}_{\UI}\q^{+}_{\UI}(\dot{z} + \ri\dot{\q}^{-}_{\UJ}\q^{+}_{\UJ})
  -\dot{\q}^{+}_{\UI}\q^{+}_{\UI}(\dot{u}^{--} - \ri\dot{\q}^{-}_{\UJ}\q^{-}_{\UJ})
  \right) + 4\ri z^{-1}\dot{\q}^{-}_{\UI}\dot{\q}^{+}_{\UI}\,. 
\end{align}
\esubeq
As a result, the $\a$ term in \eqref{3d embed model} generates purely fermionic contributions. 
Making use of the above expressions and the components of the supervielbein derived in \cite{KT}, the models \eqref{3d model} and \eqref{3d embed model} can be shown to coincide to leading order provided one fixes 
\begin{align}
    \a = -\frac{\o}{8 S^{2}}\,. 
\end{align}

\section{AdS superspaces in two dimensions} \label{section 3}

As discussed in the introduction, AdS superspaces correspond to those conformal supergravity backgrounds characterised by Lorentz-invariant and covariantly constant torsion and curvature tensors. 
In two dimensions, the $(p,q)$ conformal supergravity backgrounds were described in \cite{KR} and are reviewed in appendix \ref{appendix b}. A cursory inspection of the results of this appendix allows one to see that, since the $(p,0)$ superspaces do not possess any Lorentz-invariant torsion superfields, two-dimensional $(p,0)$ AdS superspaces cannot be obtained from this setup.\footnote{As mentioned in the introduction, other AdS superspaces based on the supergroups given in \cite{GST} are possible, but their formulations have yet to be developed.}

\subsection{$\cN$-extended AdS superspace}

To begin, we consider a general curved $(p,q)$, $p \geq q > 0\,,$ superspace \eqref{(p,q)algebra}. In order to describe AdS, it is necessary to impose the constraints
\begin{align}
    \label{6.1}
    X_{++}^{\OI \OJ} = 0~, \qquad X_{--}^{\UI \UJ} = 0~, \qquad \cD_{A} S^{\OJ \UK} = 0~,
\end{align}
where the first (second) constraint should be omitted for $p=1$ ($q=1$). There are non-trivial integrability conditions associated with the final constraint in eq. \eqref{6.1}. Specifically, by making use of eq. \eqref{(p,q)algebra}, it may be shown that
\begin{subequations}
    \label{6.2}
\begin{align}
    \d^{\OI \OK} S^{\OM \UJ} S^{\OM \UL} - \d^{\UL\UJ} S^{\OI \UM} S^{\OK \UM} = 0~,
\end{align}
which, assuming $S^{\OI \UJ} \neq 0$, is consistent only for $p=q\equiv\cN$ and implies
\begin{align}
    S^{\OI \UK} S^{\OJ \UK} = \d^{\OI \OJ} S^2~, \qquad S^{\OK \UI} S^{\OK \UJ} = \d^{\UI \UJ} S^2~, \qquad S^2 := \frac{1}{\cN} S^{\OI \UJ} S^{\OI \UJ}~.
\end{align}
\end{subequations}
We emphasise that this result indicates that two-dimensional $(p,q)$ AdS superspaces with $p \neq q$ do not exist. In keeping with \cite{KR}, we will refer to the $(\cN,\cN)$ supergeometry as $\cN$-extended AdS superspace.

In the present framework, the geometry of $\cN$-extended AdS superspace is controlled by the isotensor $S^{\OI\UJ}$. This is in contrast to the setup of \cite{KR}, where the special frame 
\begin{align}
    \label{6.4}
    S^{\OI\UJ} = \d^{\OI\UJ} S~, \qquad \cD_A S = 0~,
\end{align}
was chosen, which breaks the $R$-symmetry group down to the diagonal subgroup of $\sSO(\cN) \times \sSO(\cN)$ It should be emphasised that this gauge can only be reached if one performs a local $\sSO(\cN) \times \sSO(\cN)$ transformation, which spoils the conformal flatness of our frame.\footnote{We will say that a frame is conformally flat if it can be obtained from the flat frame by a super-Weyl transformation, see section \ref{Section6.2}.} For this reason we will not employ frame \eqref{6.4} unless otherwise stated. 

Keeping in mind constraints \eqref{6.2}, the algebra of covariant derivatives describing $\cN$-extended AdS superspace is simply
\begin{subequations} \label{2DAdSalgebra}
	\bea
	\{ \cD_{+}^{\OI}, \cD_{+}^{\OJ} \} &=& 2 \ri \d^{\OI \OJ} \cD_{++} ~, \qquad \{ \cD_{-}^{\UI}, \cD_{-}^{\UJ} \} = 2 \ri \d^{\UI \UJ} \cD_{--} ~, \\
	\{ \cD_{+}^{\OI}, \cD_{-}^{\UJ} \} &=& 4 \ri S^{\OI \UJ} M - 2 \ri S^{\OK \UJ} \mathfrak{L}^{\OK \OI} + 2 \ri S^{\OI \UK} \mathfrak{R}^{\UK \UJ} ~, 
	 \\
	\big[ \cD_{+}^{\OI} , \cD_{--} \big]
	& = & - 2 S^{\OI \UJ} \cD_{-}^{\UJ} ~, \qquad 
	\big[ \cD_{-}^{\UI} , \cD_{++} \big]
	= 2 S^{\OJ \UI} \cD_{+}^{\OJ} ~,
	\\
	\big[ \cD_{+}^{\OI} , \cD_{++} \big]
	& = & 0 ~, \qquad
	\big[ \cD_{-}^{\UI} , \cD_{--} \big] =  0  ~,
	\\
	\big[ \cD_{++} , \cD_{--} \big]
	& = & - 8  S^2 M
	~.
	\eea
\end{subequations}
The AdS curvature $S^{\OI \UJ}$ is related to the scalar curvature by $\cR = - 16 S^2 < 0$. Going to the frame \eqref{6.4} yields the $\cN$-extended AdS supergeometry of \cite{KR}.

\subsection{Conformally flat frame}
\label{Section6.2}

In this section we sketch a manifestly conformally flat realisation for the $\cN$-extended AdS superspace described above. We recall that a frame is called conformally flat if its covariant derivatives $\cD_{A}$ are related to the flat ones $D_{A}$ by a super-Weyl transformation \eqref{2d sw trf}
\begin{subequations} \label{ads frame}
\begin{align}
	{\cD}_+^\OI &= \re^{\hf \s} \Big( D_+^\OI + D_+^\OI \s M - D_+^\OJ \s \mathfrak{L}^{\OJ \OI} \Big) ~, \\
	\cD_-^\UI &= \re^{\hf \s} \Big( D_-^\UI - D_-^\UI \s M - D_-^\UJ \s \mathfrak{R}^{\UJ \UI} \Big) ~, \\
	\cal{D}_{++} &= \re^{\s} \Big( \partial_{++} - \ri D_+^\OI \s D_+^\OI + \partial_{++} \s M + \frac{\ri}{2} D_{+}^{\OI} \s D_{+}^{\OJ} \s \mathfrak{L}^{\OJ \OI} \Big) ~, \\
	\cal{D}_{--} &= \re^{\s} \Big( \partial_{--} - \ri D_-^\UI \s D_-^\UI - \partial_{--} \s M + \frac{\ri}{2} D_{-}^{\UI} \s D_{-}^{\UJ} \s \mathfrak{R}^{\UJ \UI} \Big) ~,
\end{align}
\end{subequations}
where $\s = \bar{\s}$ and the flat spinor covariant derivatives $D_{+}^{\OI}$ and $D_{-}^{\UI}$ are defined as follows
\bsubeq
\bea
D_{+}^{\OI}
:=
\frac{\pa}{\pa\q^{+ \OI}}
+ \ri \q^{+ \OI} \pa_{++}
~,\quad
D_{-}^{\UI}
:=
\frac{\pa}{\pa\q^{- \UI}}
+ \ri \q^{- \Iu}\pa_{--}
~.
\eea
Their algebra is given by
\begin{align}
 \{D_{+}^{\OI}\,,D_{+}^{\OJ}\} = 2\ri\d^{\OI\OJ}\partial_{++}\,, \qquad \{D_{-}^{\UI}\,,D_{-}^{\UJ}\} = 2\ri\d^{\UI\UJ}\partial_{--}\,, \qquad \{D_{+}^{\OI}\,,D_{-}^{\UJ}\} = 0\,.
\end{align}
\esubeq
It follows from \eqref{ads frame} that the components of the supervielbein one forms $E^{A}$ are given by
\begin{subequations} 
    \begin{align} 
        {E}^{++} &= \re^{-\s} \P^{++}~, \qquad {E}^{+ \OI} = \re^{- \frac \s 2} \Big( \rd \q^{+ \OI} + \ri D_+^{\OI} \s \P^{++} \Big) ~, \\
        {E}^{--} &= \re^{-\s} \P^{--}~, \qquad {E}^{- \UI} = \re^{- \frac \s 2} \Big( \rd\q^{- \UI} + \ri D_-^{\UI} \s \P^{--} \Big) ~,
    \end{align}
\end{subequations}
where $\P^{++} = \rd x^{++} + \ri\q^{+ \OI}\rd\q^{+ \OI}$ and $\P^{--} = \rd x^{--} + \ri\q^{- \UI}\rd\q^{- \UI}$ are the two-dimensional analogues of the Volkov-Akulov supersymmetric one-form. 

As outlined above, in order for the frame \eqref{ads frame} to correspond to AdS we must impose the constraints \eqref{6.1}. 
This leads to the following conditions on the super-Weyl parameter
\begin{subequations}
    \begin{align}
        {X}_{++}^{\OI \OJ} &= 0 \quad \implies \quad [D_+^{\OI}, D_{+}^{\OJ}] \re^{\s} = 0 ~, \\
	{X}_{--}^{\UI \UJ} &= 0 \quad \implies \quad [D_-^{\UI}, D_{-}^{\UJ}] \re^{\s} = 0 ~.
    \end{align}
\end{subequations}
Furthermore, in such a frame the torsion superfield $S^{\OI\UJ}$ is given by
\begin{align}
    S^{\OI\UJ} = \re^{\s} D_{+}^\OI D_{-}^\UJ \s\,.
\end{align}

\subsection{Interval deformation}
In two dimensions, in the lightcone coordinates used above, the standard supersymmetric interval takes the form
\begin{align}
    \rd s^{2} = E^{++}E^{--}\,. 
\end{align}
We can introduce a deformation to this interval as 
\begin{align}
    \rd s^{2} = E^{++}E^{--} + \frac{\ri\o}{S^{2}}S^{\OI\UJ}E^{+\OI}E^{-\UJ}\,,
\end{align}
for a real, dimensionless parameter $\o$. 
It then follows that there is a corresponding superparticle model, given by 
\begin{align}
    S = \frac{1}{2}\int\rd\t\frak{e}^{-1}\left\{ \dt{E}{}^{++}\dt{E}{}^{--} + \frac{\ri\o}{S^{2}}S^{\OI\UJ}\dt{E}{}^{+\OI}\dt{E}{}^{-\UJ} - (\frak{e}m)^{2}\right\}
\end{align}
 In a conformally flat frame \eqref{ads frame}, we have the following 
 \bsubeq
 \begin{align}
\dt{E}{}^{++}\dt{E}{}^{--} &= \re^{-2\s}\dt{\P}{}^{2}\,, \qquad \qquad \qquad \dt{\P}{}^{2}:= \dt{\P}{}^{++}\dt{\P}{}^{--}\,, 
\\
S^{\OI\UJ}\dt{E}{}^{+\OI}\dt{E}{}^{-\UJ} &= 
\re^{-\s}S^{\OI\UJ}\left\{ \dot{\q}^{+\OI}\dot{\q}^{-\UJ} + \ri D_{+}^{\OI}\s\dot{\q}^{-\UJ}\dt{\P}{}^{++} + \ri\dot{\q}^{+\OI}D_{-}^{\UJ}\s\dt{\P}{}^{--} 
     -D_{+}^{\OI}\s D_{-}^{\UJ}\s\dt{\P}{}^{2} \right\}\,,
 \end{align}
 \esubeq
where $\dt{\P}{}^{++} = \dot{x}^{++} + \ri\q^{+ \OI}\dot{\q}{}^{+ \OI}$ and $\dt{\P}{}^{--} = \dot{x}{}^{--} + \ri\q^{- \UI}\dot{\q}{}^{- \UI}\,.$

\section{Discussion}\label{discussion}

In this paper we have extended the recent construction of new superparticle models in the four and five-dimensional AdS superspaces given in \cite{KKR2, KK} to the two and three-dimensional cases. 
In the three-dimensional case, we derived these models from both the supergravity setting and the embedding formalism. 
To the best of our knowledge, such an embedding formalism is yet to be developed for the two-dimensional AdS supergeometry. This is an interesting problem which will be discussed elsewhere. 

The analysis of section \ref{section 2} only holds for those AdS superspaces with vanishing super-Cotton tensor $X^{IJKL} = 0\,.$ 
As described in \cite{KLTM}, there are also superspaces of the form AdS$_{(3|\N,0)}$, with $\N \geq 4$, possessing non-zero $X^{IJKL}$. 
Given the homogeneous transformation rule of $X^{IJKL}$, eq. \eqref{t trans}, such superspaces are necessarily not conformally flat. 
The presence of such a dimensionful superfield, however, allows us to construct further deformations of the AdS supersymmetric interval.
We will briefly outline this below.

It follows from eq. \eqref{x const} that $X^{KLM(I}X^{J)}{}_{KLM}$ is unconstrained. 
With the following notation, 
\bsubeq
\begin{align}
    |X| &:= \sqrt{X^{IJKL}X_{IJKL}}\,\,,
    \\
    X^{IJ}& := X^{KLM(I}X^{J)}{}_{KLM}\,, \label{sym x}
\end{align}
\esubeq
we can then introduce an additional quadratic deformation, unique to the $(\N\,,0)$ case, as
\begin{align} \label{new def}
    \rd s^{2} = \eta_{ab}E^{a}E^{b} + \left\{\frac{\o}{S}\d^{IJ} 
 + \frac{\l}{|X|^{3}}X^{IJ}\right\}\ve_{\ab}E_{I}^{\a}E_{J}^{\b} \,,
\end{align}
where both $\o$ and $\l$ are real, dimensionless parameters.
For $\N=4$, $X^{IJKL} = X\ve^{IJKL}$ and so \eqref{sym x} is proportional to $\d^{IJ}$. 
The $\o$ and $\l$ terms then correspond to the same structure, and we return to the model \eqref{3d model}. 
The story is different for $\N \geq 5\,,$ however, for which eq. \eqref{new def} contains additional structure.
The complete analysis will be provided elsewhere.

\noindent
{\bf Acknowledgements:}\\ 
We are grateful to Sergei Kuzenko for suggesting the problem and for comments on the manuscript.
We also thank Evgeny Ivanov for pointing out important references. 
The work of NEK is supported by the Australian Government Research Training Program Scholarship.

\appendix

\section{Conformal supergravity in three dimensions} \label{appendix a}

Our conventions coincide with those used in \cite{KLTM,KLTM2}.
As a starting point, we consider a three-dimensional curved superspace $\cM^{3|2\N}$ which is parametrised by local bosonic $(x^{m})$ and fermionic $(\q_{\tt I}^{\m})$ coordinates 
\begin{align}
    z^{M} = (x^{m}\,, \q_{\tt I}^{\m})\,, \qquad m = 0\,,1\,,2\,, \quad \m = 1\,,2\,, \quad {\tt I} = 1\,, \ldots\,, \N\,.
\end{align}
Its structure group is $\sSL(2\,,\mathbb{R}) \times \sSO(\N)\,,$ and so the covariant derivatives take the form 
\begin{align}
    \cD_{A} \equiv (\cD_{a}\,, \cD_{\a}^{I}) = E_{A} + \O_{A} + \F_{A}\,. 
\end{align}
Here, $E_{A} = E_{A}{}^{M}\partial_{M}$ denotes the frame field, with $E_A{}^M$ being the inverse supervielbein.\footnote{The supermatrix $E_{A}{}^M$ is assumed to be non-singular, hence there exists a unique inverse $E_M{}^A$. It defines the supervielbein one forms $E^A = \rd z^M E_M{}^A$, which constitute a basis for the cotangent space at each point.}
The superfield 
\begin{align}
    \O_{A} = \frac{1}{2}\O_{A}{}^{bc}M_{bc} = - \O_{A}{}^{b}M_{b} =  \frac{1}{2}\O_{A}{}^{\g\d}M_{\g\d}\,, \qquad M_{ab} = - M_{ba}\,, \quad M_{\ab} = M_{\b\a}\,,
\end{align}
denotes the Lorentz connection; and 
\begin{align}
    \F_{A} = \frac{1}{2}\F_{A}{}^{KL}\N_{KL}\,, \qquad \N_{KL} = -\N_{LK}\,,
\end{align}
is the $\sSO(\N)$ connection. 

Our spinor conventions in three dimensions follow \cite{KLTM2}, and are compatible with the 4D two-component formalism used in \cite{WB,BK}.
In particular, two-component spinor indices are raised and lowered with the $\sSL(2;\mathbb{R})$ invariant tensors 
\begin{align}
    \ve^{\ab} = \left(\begin{array}{cc}
     ~0~    & 1 \\
       -1  & ~0~
    \end{array}\right)\,, \qquad \ve_{\ab} = \left(\begin{array}{cc}
  ~0~       & -1 \\
      1   & ~0~
    \end{array}\right)\,,
\end{align}
according to the rule 
\begin{align}
    \psi_{\a} = \ve_{\ab}\psi^{\b}\,, \qquad \psi^{\a} = \ve^{\ab}\psi_{\b}\,.
\end{align}
Furthermore, the Lorentz generators with spinor indices are related to those with vector indices by the rule 
\begin{align}
    M_{\ab} = \frac{1}{2}(\g^{a})_{\ab}\ve_{abc}M^{bc}\,,
\end{align}
where $(\g^{a})_{\ab} = (\id_{2}\,, \s_{1}\,, \s_{3})$ are the gamma-matrices in three dimensions. 

The generators of $\sSL(2\,,{\mathbb{R}}) \times \sSO(\N)$ act on the covariant derivatives as follows:
%
\begin{align}
    [M_{\ab}\,, \cD_{\g}^{I}] &= \ve_{\g(\a}\cD_{\b)}^{I}\,, \quad [M_{ab}\,, \cD_{c}] = 2\eta_{c[a}\cD_{b]}\,, \quad 
    [\N_{KL}\,, \cD_{\a}^{I}] = 2\d^{I}_{[K}\cD_{\a L]}\,,
\end{align}
%
and amongst themselves as 
\bsubeq
\begin{align}
    [M_{ab}\,, M_{cd}] &= \eta_{ad}M_{bc} - \eta_{ac}M_{bd} + \eta_{bc}M_{ad} - \eta_{bd}M_{ac}\,,
    \\
    [M_{\ab}\,, M_{\g\d}] &= \frac{1}{2}\left( \ve_{\a\g}M_{\b\d}+\ve_{\a\d}M_{\b\g} + \ve_{\b\g}M_{\a\d} + \ve_{\b\d}M_{\a\g} \right) \,,
    \\
    [\N_{IJ}\,,\N_{KL}] &= \d_{IL}\N_{JK} - \d_{IK}\N_{JL} + \d_{JK}\N_{IL} - \d_{JL}\N_{IK} \,. 
\end{align}
\esubeq
All other commutators vanish. 

The supergravity gauge group is generated by local transformations of the form 
\begin{align}
    \d_{\cK}\cD_{A} = [\cK\,,\cD_{A}]\,, \qquad \cK = \x^{C}\cD_{C} + \frac{1}{2} K^{cd}M_{cd} + \frac{1}{2}\L^{PQ}\N_{PQ}\,,
\end{align}
with the parameters obeying natural reality conditions. 
These transformations act on tensor superfields (with indices suppressed) as 
\begin{align}
    \d_{\cK}\cU = \cK\cU\,. 
\end{align}

The structure of the algebra of covariant derivatives differs between the $\N=1$ and $\N>1$ cases.
We specify both cases below. 

\subsection{$\N=1$}

For $\N=1$, the algebra of covariant derivatives is given by 
\bsubeq\label{n=1 alg}
\begin{align}
\{\cD_{\a}\,,\cD_{\b}\} &= 
2\ri\cD_{\ab}-4\ri S M_{\ab}\,,
\\
[\cD_{\ab}\,,\cD_{\g}] &= 
-2S\ve_{\g(\a}\cD_{\b)} + 2\ve_{\g(\a}C_{\b)\d\r}M^{\d\r} + \frac{2}{3}\left( (\cD_{\g}S)M_{\ab} - 4(\cD_{(\a}S)M_{\b)\g} \right)\,,
\\
[\cD_{a}\,,\cD_{b}]&= 
\frac{1}{2}\ve_{abc}(\g^{c})^{\ab}\bigg\{ -\ri C_{\ab\g}\cD^{\g} - \frac{4\ri}{3}(\cD_{\a}S)\cD_{\b} + \ri\cD_{(\a}C_{\b\g\d)}M^{\g\d} 
\\
& \qquad \qquad \qquad ~~~~ -\left(\frac{2\ri}{3}(\cD^{2}S)+4S^{2}\right)M_{\ab}\bigg\}\,, 
\end{align}
\esubeq
see \cite{KLTM2}. 
It is expressed in terms of the dimension-1 scalar $S$ and a totally symmetric dimension-$3/2$ spinor $C_{\a\b\g} = C_{(\a\b\g)}\,,$ obeying the constraint 
\begin{align}
    \cD_{\a}C_{\b\g\d} = \cD_{(\a}C_{\b\g\d)} - \ri\ve_{\a(\b}\cD_{\g\d)}S\,. 
\end{align}

The structure of the algebra \eqref{n=1 alg} is invariant under the super-Weyl transformations 
\bsubeq \label{n=1 sw cd}
\begin{align}
  \hat{\cD}_{\a} &= \re^{\frac{1}{2}\s}\left\{ \cD_{\a} + (\cD^{\b}\s)\cM_{\ab} \right\}\,,
    \\
    \hat{\cD_{a}} &= \re^{\s}\bigg\{\cD_{a} + \frac{\ri}{2}(\g_{a})^{\ab}(\cD_{\a}\s)\cD_{\b} + \ve_{abc}(\cD^{b}\s)\cM^{c}  - \frac{\ri}{8}(\g_{a})^{\ab}(\cD^{\r}\s)(\cD_{\r}\s)\cM_{\ab} 
    \bigg\}\,. 
\end{align}
\esubeq
The corresponding transformations of the torsion superfields are 
\bsubeq
\begin{align}
    \hat{S} &= \frac{\ri}{2}\re^{\frac{3}{2}\s}\left\{\cD^{2} - 2\ri S\right\}\re^{-\frac{1}{2}\s}
    \,,
    \\
    \hat{C}_{\a\b\g} &= -\frac{1}{2}\re^{\frac{1}{2}\s}\left\{D_{(\ab}D_{\g)} -2C_{\a\b\g} \right\}\re^{\s}\,. 
\end{align}
\esubeq
The finite form of these super-Weyl transformations was given in \cite{KNTM}. 
It follows from \eqref{n=1 sw cd} that the transformations for the supervielbein one forms are given by
\bsubeq 
\begin{align}
    \hat{E}^{a} &= \re^{-\s}E^{a}\,, 
    \\
    \hat{E}^{\a} &= \re^{-\frac{1}{2}\s}\left\{ E^{\a} - \frac{\ri}{2}(\cD_{\b}\s)E^{\ab} \right\}\,. 
\end{align}
\esubeq

\subsection{$\N>1$}
We now detail the $\N>1$ case.
Up to dimension-1, the covariant derivatives $\cD_{A}$ obey the following algebra
\begin{align}
\label{3d alg}
    \{\cD_{\a}^{I}\,, \cD_{\b}^{J}\} &= 2\ri\d^{IJ}\cD_{\ab} - 2\ri\ve_{\ab}C^{\g\d IJ}M_{\g\d} - 4\ri S^{IJ}M_{\ab} \\
    \notag
     & \quad + \left(\ri \ve_{\ab} X^{IJKL} - 4\ri\ve_{\ab}S^{K[I}\d^{J]L} + \ri C_{\ab}{}^{KL}\d^{IJ} - 4\ri C_{\ab}{}^{K(I}\d^{J)L}   \right)\N_{KL}\,.
\end{align}
This algebra is given in terms of the real superfields $X^{IJKL}$, $S^{IJ}$ and $C_{a}{}^{IJ}$ which have the following symmetry properties 
\begin{align} \label{3d torsions}
    X^{IJKL} = X^{[IJKL]}\,, \qquad S^{IJ} = S^{(IJ)}\,, \qquad C_{a}{}^{IJ} = C_{a}{}^{[IJ]}\,.
\end{align}
Importantly, $C_{a}{}^{IJ} = 0$ for $\N=1$, and $X^{IJKL} = 0$ for $\N < 4\,.$

The structure of the algebra \eqref{3d alg} is invariant under super-Weyl transformations. The super-Weyl transformations of the covariant derivatives are given by 
\bsubeq \label{cd trans}
\begin{align}
    \hat{\cD}_{\a}^{I} &= \re^{\frac{1}{2}\s}\left\{ \cD_{\a}^{I} + (\cD^{\b I}\s)\cM_{\ab} + (\cD_{\a J}\s)\N^{IJ} \right\}\,,
    \\
    \hat{\cD_{a}} &= \re^{\s}\bigg\{\cD_{a} + \frac{\ri}{2}(\g_{a})^{\ab}(\cD^{K}_{\a}\s)\cD_{\b K} + \ve_{abc}(\cD^{b}\s)\cM^{c} + \frac{\ri}{16}(\g_{a})^{\ab}([\cD^{K}_{\a}\,,\cD^{L}_{\b}]\s)\N_{KL}
    \\
    \notag 
    & ~~~~~~~\qquad - \frac{\ri}{8}(\g_{a})^{\ab}(\cD_{K}^{\r}\s)(\cD^{K}_{\r}\s)\cM_{\ab} + \frac{3\ri}{8}(\g_{a})^{\ab}(\cD^{K}_{\a}\s)(\cD_{\b}^{L}\s)\N_{KL}
    \bigg\}\,. 
\end{align}
\esubeq
The dimension-1 torsion superfields transform as
\bsubeq \label{t trans}
\begin{align}
    \hat{S}^{IJ} &= \re^{\s}\left\{ S^{IJ} - \frac{\ri}{8}([\cD^{\r I}\,,\cD_{\r}^{J}]\s) + \frac{\ri}{4}(\cD^{\r I}\s)(\cD_{\r}^{J}\s) - \frac{\ri}{8}\d^{IJ}(\cD^{\r}_{K}\s)(\cD^{K}_{\r}\s)  \right\}\,,
    \\
    \hat{C}_{a}{}^{IJ} &= \re^{\s}\left\{  C_{a}{}^{IJ} - \frac{\ri}{8}(\g_{a})^{\ab}([\cD_{\a}^{I}\,,\cD_{\b}^{J}]\s) - \frac{\ri}{4}(\g_{a})^{\ab}(\cD_{\a}^{I}\s)(\cD_{\b}^{J}\s) \right\}\,,
    \\
    \hat{X}^{IJKL} &= \re^{\s}X^{IJKL}\,. 
\end{align}
\esubeq
It follows from \eqref{cd trans} that the corresponding transformations for the supervielbein one forms $E^{A}$ are 
\bsubeq \label{vb trans}
\begin{align}
    \hat{E}^{a} &= \re^{-\s}E^{a}\,, 
    \\
    \hat{E}_{I}^{\a} &= \re^{-\frac{1}{2}\s}\left\{ E_{I}^{\a} - \frac{\ri}{2}(\cD_{\b I}\s)E^{\ab} \right\}\,. 
\end{align}
\esubeq
The relations \eqref{cd trans}, \eqref{t trans}, and \eqref{vb trans} are used in the main body to define a conformally flat frame for AdS$_{(3|p\,,q)}\,.$

\section{Conformal $(p,q)$ supergravity in two dimensions} \label{appendix b}

We consider a two-dimensional curved $(p,q)$ superspace $\cM^{(2|p,q)}$ parametrised by local coordinates $z^M = (x^{\hat + \hat +},x^{\hat - \hat -},\q^{\hat{+}\overline{\cI}},\q^{\hat{-}\underline{\cI}})$ where $\overline{\cI} = 1, \dots , p$ and $\underline{\cI} = 1, \dots , q$.\footnote{For convenience we will always assume $p \geq q$.} Its structure group is chosen to be\footnote{We recall that $\sSO(1,1)$ is the two-dimensional Lorentz group, while $\sSO(p)\times\sSO(q)$ is the $R$-symmmetry group of $(p,q)$ supersymmetry.} $\sSO(1,1) \times \sSO(p) \times \sSO(q)$ and so the curved covariant derivatives $\cD_A = (\cD_{++}, \cD_{--}, \cD_{+}^{\overline{I}}, \cD_{-}^{\underline{I}})$ take the form
\begin{align}
    \cD_A = E_A + \O_A M +  \hf \F_A{}^{\overline{J} \overline{K}} \frak{L}^{\overline{J} \overline{K}} + \hf \F_A{}^{\underline{J} \underline{K}} \frak{R}^{\underline{J}\underline{K}}~.
\end{align}
Here, $E_A = E_A{}^M \partial_M$ denotes the frame field, with $E_A{}^M$ being the inverse supervielbein.\footnote{As before, the supermatrix $E_{A}{}^M$ is assumed to be non-singular.} The superfield $\O_A$ denotes the Lorentz connection, while $\F_A{}^{\overline{J} \overline{K}}$ and $\F_A{}^{\underline{J} \underline{K}}$ are associated with the left and right and $R$-symmetry groups, respectively. 

The Lorentz generator $M$ acts on the covariant derivatives $\cD_A$ as follows
\begin{subequations}
    \begin{align}
        [M,\cD_{++}] &= \cD_{++} ~, \qquad [M,\cD_{+}^{\overline{I}}] = \hf \cD_{+}^{\overline{I}}~, \\
        [M,\cD_{--}] &= - \cD_{--} ~, \qquad [M,\cD_{-}^{\underline{I}}] = - \hf \cD_{-}^{\underline{I}}~.
    \end{align}
\end{subequations}
The $R$-symmetry generators $\mathfrak{L}^{\OI \OJ}$ and $\mathfrak{R}^{\UI \UJ}$ act amongst themselves in accordance with
\begin{subequations}
\begin{align}
        [\mathfrak{L}^{\OI \OJ} , \mathfrak{L}^{\OK \OL}] &= 2 \d^{\OK[\OI} \mathfrak{L}^{\OJ] \OL} - 2 \d^{\OL[\OI} \mathfrak{L}^{\OJ] \OK}~, \\
        [\mathfrak{R}^{\UI \UJ} , \mathfrak{R}^{\UK \UL}] &= 2 \d^{\UK[\UI} \mathfrak{R}^{\UJ] \UL} - 2 \d^{\UL[\UI} \mathfrak{R}^{\UJ] \UK}~.
\end{align}
and on the covariant derivatives as follows
\begin{align}
        [\mathfrak{L}^{\OI \OJ} , \cD_{+}^{\OK}] = 2 \d^{\OK [\OI} \cD_{+}^{\OJ]}~, \qquad 
        [\mathfrak{R}^{\UI \UJ} , \cD_{-}^{\UK}] = 2 \d^{\UK [\UI} \cD_{-}^{\UJ]}~,
\end{align}
\end{subequations}
where all other commutators vanish.

The supergravity gauge freedom of this geometry includes local $\mathcal{K}$-transformations of the form
\bea
    \label{SGtransformations}
    \delta_{\mathcal K} \cD_A &=& [\mathcal{K},\cD_A] ~, \qquad
    \mathcal{K} = \xi^B \cD_B + K M +\hf \r^{\OI \OJ} \mathfrak{L}^{\OI \OJ}
	+ \hf \r^{\UI \UJ} \mathfrak{R}^{\UI \UJ} ~,
\eea
which act on tensor superfields $\mathcal{U}$ (with its indices suppressed) as
\begin{align}
	\d_\cK \cU = \cK \cU ~.
\end{align}

The covariant derivatives $\cD_A$ obey graded commutation relations
\begin{align}
    [\cD_A,\cD_B] = \cT_{AB}{}^{C} \cD_C + \cR_{AB} M + \hf \cF_{AB}{}^{\OK \OL} \mathfrak{L}^{\OK \OL} + \hf \cF_{AB}{}^{\UK \UL} \mathfrak{R}^{\UK \UL}~,
\end{align}
where $\cT_{AB}{}^{C}$ denotes the torsion, $\cR_{AB}$ is the Lorentz curvature, and the pair $\cF_{AB}{}^{\OK \OL}$ and $\cF_{AB}{}^{\UK \UL}$ constitute the $R$-symmetry curvature. To describe conformal supergravity, these tensors must obey certain covariant constraints. For general $p,q\geq0$, these constraints and their solution were given in \cite{KR}. We emphasise that the constraints differ functionally between the cases $q=0$ and $q>0$, hence we will describe the corresponding geometries separately.


\subsection{Curved $(p,0)$ superspaces}

First, we will study the case of curved $(p,0)$ supergeometries, which are characterised by $q=0$. We emphasise that in this case the spinor derivatives $\cD_{-}^{\UI}$ and corresponding $R$-symmetry generators $\mathfrak{R}^{\UI \UJ}$ are absent. In this case the algebra of covariant derivatives takes the form
\begin{subequations} \label{(p,0)algebra}
	\bea
	\{ \cD_{+}^{\OI}, \cD_{+}^{\OJ} \} &=& 2 \ri \d^{\OI \OJ} \cD_{++} + 4 \ri X_{++}^{\OK (\OI} \mathfrak{L}^{\OJ) \OK} ~, \\
	\big[ \cD_{+}^{\OI} , \cD_{--} \big]
	& = & 4 \ri G_-^\OI M - 2 \ri G_-^\OJ \mathfrak{L}^{\OJ \OI}
	~, \\
        \label{6.7c}
	\big[ \cD_{+}^{\OI} , \cD_{++} \big]
	& = & - 2 X_{++}^{\OI \OJ} \cD_+^{\OJ} - \frac{2}{p-1} \cD_+^\OJ X_{++}^{\OJ \OK} \mathfrak{L}^{\OK \OI} ~,
	\\
	\big[ \cD_{++} , \cD_{--} \big]
	& = & 2 G_-^\OI \cD_+^\OI - \frac{4}{p} \cD_{+}^\OI G_{-}^\OI M ~.
	\eea
\end{subequations}
Hence, the geometry of this superspace is controlled by the torsions $X_{++}^{\OI \OJ} = - X_{++}^{\OJ \OI}$ and $G_{-}^\OI$,\footnote{Note that, for $p=1$, the $\sSO(p)$ curvature is absent, and so $G_-$ is the only torsion superfield. Hence, the apparent singularity in eq. \eqref{6.7c} is artificial.} which obey the Bianchi identities
\begin{align}
	\label{5.24}
	\cD_+^{\OI} X_{++}^{\OJ \OK} = \frac{2}{p-1} \d^{\OI [\OJ} \cD_+^{|\OL} X_{++}^{\OL| \OK]} ~, \qquad
	\cD_+^{\OI} G_-^{\OJ} = \frac{1}{p} \d^{\OI \OJ} \cD_+^{\OK} G_-^{\OK} + \cD_{--} X_{++}^{\OI \OJ} ~.
\end{align}

In the $p=1$ case, supergeometry \eqref{(p,0)algebra} was originally constructed in \cite{BMG,EO,GGMT}. For $p\geq2$, the corresponding supergeometries were described in \cite{EO}, though the $p=2$ case appeared earlier \cite{BG}.\footnote{The structure group for $(2,0)$ supergravity was enlarged from $\sSO(1,1)$ to $\sSO(1,1) \times \sU(1)$ in \cite{GO}.} The relationship between the present supergeometry and the $(2,0)$ supergeometry of \cite{BG} was described in \cite{KR}.

In order for this superspace geometry to describe $(p,0)$ conformal supergravity, the supergravity gauge group should include super-Weyl transformations, which are characterised by the property that they are local rescalings preserving the structure of the algebra \eqref{(p,0)algebra}. These prove to be parametrised by a real scalar superfield $\s = \bar{\s}$ and take the form:
\begin{subequations}
	\label{(p,0)SW}
	\begin{align}
		\hat{\cD}_+^\OI &= \re^{\hf \s} \Big( \cD_+^\OI + \cD_+^\OI \s M - \cD_+^\OJ \s \mathfrak{L}^{\OJ \OI} \Big)~, \\
		\hat{\cD}_{++} &= \re^{\s} \Big( \cD_{++} - \ri \cD_+^\OI \s \cD_+^\OI + \cD_{++} \s M + \frac{\ri}{2} \cD_{+}^{\OI} \s \cD_{+}^{\OJ} \s \mathfrak{L}^{\OJ \OI} \Big) ~, \\
		\hat{\cD}_{--} &= \re^{\s} \Big( \cD_{--} - \cD_{--} \s M \Big) ~, \\
		\hat{X}_{++}^{\OI \OJ} &= \re^{\s} \Big( X_{++}^{\OI \OJ} + \frac \ri 4 \big [ \cD_+^\OI , \cD_+^\OJ \big ] \s + \frac{\ri}{2} \cD_{+}^{\OI} \s \cD_{+}^{\OJ} \s \Big)~, \\
		\hat{G}_{-}^I &= \re^{\frac{3}2 \s} \Big( G_{-}^\OI + \frac \ri 2 \cD_+^\OI \cD_{--} \s \Big)~.
	\end{align}
\end{subequations}
For completeness, we also provide the corresponding transformations for the supervielbein one forms $E^A$:
\begin{subequations}
    \begin{align}
        \hat{E}^{++} &= \re^{-\s} E^{++}~, \qquad \hat{E}^{+ \OI} = \re^{- \frac \s 2} \Big( E^{+ \OI} + \ri \cD_+^{\OI} \s E^{++} \Big) ~, \\
        \hat{E}^{--} &= \re^{-\s} E^{--}~. 
    \end{align}
\end{subequations}

\subsection{Curved $(p,q)$ superspaces}

In this subsection we will consider the remaining case of curved $(p,q)$ supergeometries wherein $p \geq q >0$. It was shown in \cite{KR} that the algebra of covariant derivatives for this supergeometry take the form
\begin{subequations} \label{(p,q)algebra}
	\bea
	\{ \cD_{+}^{\OI}, \cD_{+}^{\OJ} \} &=& 2 \ri \d^{\OI \OJ} \cD_{++} + 4 \ri X_{++}^{\OK (\OI} \mathfrak{L}^{\OJ) \OK} ~, \\
	\{ \cD_{+}^{\OI}, \cD_{-}^{\UJ} \} &=& 4 \ri S^{\OI \UJ} M - 2 \ri S^{\OK \UJ} \mathfrak{L}^{\OK \OI} + 2 \ri S^{\OI \UK} \mathfrak{R}^{\UK \UJ} ~, \label{(p,q)algebra.b}\\
	\{ \cD_{-}^{\UI}, \cD_{-}^{\UJ} \} &=& 2 \ri \d^{\UI \UJ} \cD_{--} + 4 \ri X_{--}^{\UK (\UI} \mathfrak{R}^{\UJ) \UK} ~, \\
	\big[ \cD_{+}^{\OI} , \cD_{--} \big]
	& = & - 2 S^{\OI \UJ} \cD_{-}^{\UJ} - \frac{4}{q} \cD_{-}^{\UJ} S^{\OI \UJ} M + \frac{2}{q} \cD_{-}^{\UK} S^{\OJ \UK} \mathfrak{L}^{\OJ \OI}
	~, \\
	\big[ \cD_{-}^{\UI} , \cD_{++} \big]
	& = & 2 S^{\OJ \UI} \cD_{+}^{\OJ} - \frac{4}{p} \cD_{+}^{\OJ} S^{\OJ \UI} M - \frac{2}{p} \cD_{+}^{\OK} S^{\OK \UJ} \mathfrak{R}^{\UJ \UI} ~,
	\\
	\big[ \cD_{+}^{\OI} , \cD_{++} \big]
	& = & - 2 X_{++}^{\OI \OJ} \cD_+^{\OJ} - \frac{2}{p-1} \cD_+^\OJ X_{++}^{\OJ \OK} \mathfrak{L}^{\OK \OI} ~,
	\\
	\big[ \cD_{-}^{\UI} , \cD_{--} \big]
	& = & - 2 X_{--}^{\UI \UJ} \cD_-^{\UJ} - \frac{2}{q-1} \cD_-^\UJ X_{--}^{\UJ \UK} \mathfrak{R}^{\UK \UI}  ~,
	\\
	\big[ \cD_{++} , \cD_{--} \big]
	& = & \frac{2 \ri}{p} \cD_+^{\OJ} S^{\OJ \UI} \cD_-^{\UI} + \frac{2 \ri}{q} \cD_-^{\UJ} S^{\OI \UJ} \cD_+^{\OI} \non \\
	&& + \frac{2 \ri}{pq} \big( \big[\cD_+^{\OI}, \cD_-^{\UJ}\big] + 2(p+q) \ri S^{\OI \UJ} \big) S^{\OI \UJ} M
	~.
	\eea
\end{subequations}
We see that the geometry of this superspace is controlled by the three torsion superfields $X_{++}^{\OI \OJ} = - X_{++}^{\OJ \OI}$, $X_{--}^{\UI \UJ} = - X_{--}^{\UJ \UI}$, and $S^{\OI \UJ}$.\footnote{We note that for $p=1$ $(q=1)$ the superfield $X_{++}^{\OI \OJ}$ ($X_{--}^{\UI \UJ}$) identically vanishes. Hence, the apparent singularities in eq. \eqref{(p,q)algebra} are artificial.} These obey the differential constraints
\begin{subequations} \label{(p,q)Bianchi}
\begin{align}
	\cD_+^{\OI} S^{\OJ \UK} &= \frac 1 p \d^{\OI \OJ} \cD_+^{\OL} S^{\OL \UK} + \cD_-^\OK X_{++}^{\OI \OJ} ~, \quad 
	\cD_-^{\UI} S^{\OJ \UK} = \frac 1 q \d^{\UI \UK} \cD_-^{\UL} S^{\OJ \UL} - \cD_+^\OJ X_{--}^{\UI \UK} ~, \\
	\cD_+^{\OI} X_{++}^{\OJ \OK} &= \frac{2}{p-1} \d^{\OI [\OJ} \cD_+^{|\OL} X_{++}^{\OL| \OK]} ~, \qquad \quad 
	\cD_-^{\UI} X_{--}^{\UJ \UK} = \frac{2}{q-1} \d^{\UI [\UJ} \cD_-^{|\UL} X_{--}^{\UL| \UK]} ~.
\end{align}
\end{subequations}

In the $p=q=1$ case, this superspace geometry was originally constructed in \cite{Howe,BG}, see also \cite{Martinec,GN,RvanNZ}. For $p=q=2$ equivalent supergeometry to the present one was formulated in the works \cite{HP,GW1,GW2,GGW}, see \cite{KR} for the technical details regarding this relationship. Additionally, an alternative $(4,4)$ supergeometry with local structure group $\sSL(2,\mathbb{R}) \times \sSU(2) \times \sSU(2)$ was proposed in \cite{GTM}.\footnote{It should also be pointed out that the formulation for $(4,4)$ matter-coupled supergravity in $\sSU(2)\times\sSU(2)$ harmonic superspace was constructed in \cite{BI}.}

As was the case for the $(p,0)$ supergeometry studied above, the algebra of covariant derivatives for $(p,q)$ superspace geometries \eqref{(p,q)algebra} prove to be preserved under super-Weyl transformations parametrised by a real scalar superfield $\s = \bar{\s}$. These local rescalings are:
\begin{subequations} \label{2d sw trf}
\begin{align}
	\hat{\cD}_+^\OI &= \re^{\hf \s} \Big( \cD_+^\OI + \cD_+^\OI \s M - \cD_+^\OJ \s \mathfrak{L}^{\OJ \OI} \Big) ~, \\
	\hat{\cD}_-^\UI &= \re^{\hf \s} \Big( \cD_-^\UI - \cD_-^\UI \s M - \cD_-^\UJ \s \mathfrak{R}^{\UJ \UI} \Big) ~, \\
	\hat{\cD}_{++} &= \re^{\s} \Big( \cD_{++} - \ri \cD_+^\OI \s \cD_+^\OI + \cD_{++} \s M + \frac{\ri}{2} \cD_{+}^{\OI} \s \cD_{+}^{\OJ} \s \mathfrak{L}^{\OJ \OI} \Big) ~, \\
	\hat{\cD}_{--} &= \re^{\s} \Big( \cD_{--} - \ri \cD_-^\UI \s \cD_-^\UI - \cD_{--} \s M + \frac{\ri}{2} \cD_{-}^{\UI} \s \cD_{-}^{\UJ} \s \mathfrak{R}^{\UJ \UI} \Big) ~, \\
	\hat{X}_{++}^{\OI \OJ} &= \re^{\s} \Big( X_{++}^{\OI \OJ} + \frac \ri 4 \big[\cD_+^\OI , \cD_+^\OJ \big]  \s + \frac{\ri}{2} \cD_{+}^{\OI} \s \cD_{+}^{\OJ} \s \Big) ~, \\
	\hat{X}_{--}^{\UI \UJ} &= \re^{\s} \Big( X_{--}^{\UI \UJ} + \frac \ri 4 \big[\cD_-^\UI , \cD_-^\UJ \big] \s + \frac \ri 2 \cD_-^\UI \s  \cD_-^\UJ \s \Big) ~, \\
        \hat{S}^{\OI \UJ} &= \re^{\s} \Big( S^{\OI \UJ} + \frac \ri 2 \cD_+^\OI \cD_-^\UJ \s \Big) ~.
\end{align}
\end{subequations}
The corresponding super-Weyl transformations of the supervielbein one forms $E^A$ are as follows:
\begin{subequations}
    \begin{align}
        \hat{E}^{++} &= \re^{-\s} E^{++}~, \qquad \hat{E}^{+ \OI} = \re^{- \frac \s 2} \Big( E^{+ \OI} + \ri \cD_+^{\OI} \s E^{++} \Big) ~, \\
        \hat{E}^{--} &= \re^{-\s} E^{--}~, \qquad \hat{E}^{- \UI} = \re^{- \frac \s 2} \Big( E^{- \UI} + \ri \cD_-^{\UI} \s E^{--} \Big) ~.
    \end{align}
\end{subequations}
For the $(1,1)$ and $(2,2)$ cases, the above super-Weyl transformations reduce to those of \cite{Howe} and \cite{HP}, respectively. Additionally, it is expected that in the $(4,4)$ case the corresponding super-Weyl transformations are equivalent to those given in \cite{GTM}.

\begin{footnotesize}

\providecommand{\href}[2]{#2}\begingroup\raggedright\endgroup

\end{footnotesize}

\end{document}